\documentclass[aps,prl,preprintnumbers,showpacs,twocolumn,nofootinbib]{revtex4}
\usepackage{graphicx}
\usepackage{equation}

\newcommand{\PRE}[1]{}       

\def\gappeq{\mathrel{\rlap {\raise.5ex\hbox{$>$}}
            {\lower.5ex\hbox{$\sim$}}}}
\def\lappeq{\mathrel{\rlap{\raise.5ex\hbox{$<$}}
            {\lower.5ex\hbox{$\sim$}}}}

\def\g{\gamma}

\def\Ou{\mathcal{O}_{\mathcal{U}}}
\def\Lu{\Lambda_{\mathcal{U}}}
\def\dU{d_{\mathcal{U}}}
\def\G{\Gamma}

\begin{document}

\preprint{HIP-2007-65/TH}
\renewcommand{\thefootnote}{\fnsymbol{footnote}}

\title{
\PRE{\vspace*{2.0in}}
 Disentangling the Unparticles with polarized beams at $e^+e^-$ colliders
\PRE{\vspace*{.4in}}
}

\author{Katri Huitu$^1$ and Santosh Kumar Rai$^1$}\PRE{\vspace{.1in}}
\affiliation{{}$^1$  High Energy Physics Division, Department 
of Physics, University of Helsinki,\\ and Helsinki Institute 
of Physics, P.O. Box 64, FIN-00014 University of Helsinki, Finland \\
 }
\begin{abstract}
Recently proposed idea of unparticles arising due to a scale invariant sector
in the theory can give rise to effective operators with different Lorentz 
structures. We show that by using the different polarization 
options at the future linear $e^+ e^-$ colliders, the nature of these effective 
operators can be easily understood. The unique feature of a complex phase
in the propagator of the unparticle can also be understood uniquely for
the different spins by exploiting the initial beam polarizations at the 
International Linear Collider (ILC).
\end{abstract}

\maketitle

The recently proposed idea of the consequences of having a scale invariant
sector in a theory \cite{georgi1,georgi2}, which decouples from the low
energy theory, has drawn enormous attention due to its rich phenomenology.
Such a sector of scale invariant physics existing at a much higher scale 
can co-exist and still remain sterile to interactions at very low energies 
to hide its existence. Motivated by the Banks-Zaks theory \cite{banks}, it was
argued however, that a scale invariant sector with a nontrivial infrared
fixed-point leads to peculiar low energy behavior in an effective field theory
approach, with non-matter stuff coined as ``unparticles" ($\mathcal{U}$). 
The production of these unparticles, both on-shell and off-shell, 
can lead to interesting signatures at future accelerator experiments. 
Real production of the unparticles will lead to missing energy 
signatures. Another peculiar feature of the 
unparticles was identified as a complex ``phase" originating in the 
propagators of the unparticles \cite{georgi2,kingman1} for time like 
momenta. These phases can be a crucial test in isolating signatures for 
unparticles at future experiments, and features related to the 
phases were shown to be quite interesting \cite{georgi2,kingman1,kingman2}. 

A lot of work already exists in the literature exploring the various 
possibilities for unparticles and their signatures beyond the 
Standard Model (SM) physics \cite{unplit}. 
We focus on the importance of the complex phase that
appears as an unique feature of the unparticle theory in our analysis. 
In this work, we discuss how effectively the use of polarization of
initial beams at the proposed International Linear Collider (ILC) \cite{ilc}
can be used to isolate signatures for unparticles by looking at a few kinematic
distributions. We will focus on the simplest of scattering processes at
the ILC, {\it viz.} $e^+e^-\to f \bar{f}$ for the analysis. The different
chiralities for the fermions involved in the scattering would show quite
distinct features and this was highlighted in Ref.\cite{kingman2} for the
process $e^+e^-\to\mu^+\mu^-$ for LEP energies. The effect of the complex
phase in the propagator was shown to effect the cross-sections for
different center-of-mass energies.  Identifying such a feature at
experiments would prove quite conclusive. We show that this feature can be
very easily captured by the use of transverse beam polarizations at the
ILC. It has been shown that transverse beam polarizations can be a very useful
tool at ILC to study CP-odd observables and effects of 
beyond the SM physics \cite{budny,olsen,hikasa,rizzo,bartl}. 
We show that not only is the transverse polarization effective in
distinguishing the various effective operators of the unparticle sector,
the dependence of the cross-section on the mass of the final states also
plays a crucial role in this distinction.

We choose the following notation for introducing the beam polarizations 
\begin{eqnarray} \label{polsum}
\sum_{s_1} \bar{u}(k_1,s_1) u(k_1,s_1) 
= \frac{1}{2}(1 + P_L \g_5 + \g_5 P_T t\!\!/_1) k\!\!\!/_1 \nonumber \\ 
\sum_{s_2} \bar{v}(k_2,s_2) v(k_2,s_2) 
= \frac{1}{2}(1 - P_L \g_5 + \g_5 P_T t\!\!/_2 ) k\!\!\!/_2 
\end{eqnarray}
where $t_{1,2}$ are the transverse beam polarization 4-vectors of the
electron and positron beams, respectively. In the above equation, $P_L$ and
$P_T$ represent the degree of the longitudinal and transverse beam
polarizations. For our analysis, we chose
$|P_{L,T}^{e^-}|=0.8$ and $|P_{L,T}^{e^+}|=0.6$. For the transverse 
beam polarization 4-vectors we assume $t_1^\mu=(0,1,0,0)=-t_2^\mu$ 
(the results still hold for other choices). 
The scattering process we consider is mediated by the $\g$ and 
$Z$-boson propagators in the SM, while the ``unparticle" 
effects will show up due to the s-channel contribution coming from the 
effective unparticle operators $\Ou$ respecting the SM gauge symmetry. 
We focus on three different operators {\it viz.} the scalar, vector and 
tensor operators given respectively by
\begin{center}
\begin{eqnarray} \label{operators}
&&\lambda_0 \frac{1}{{\Lu}^{\dU - 1}} \bar{f} f \Ou \\
&&\lambda_1 \frac{1}{{\Lu}^{\dU - 1}} \bar{f} \g_\mu (1+\g_5)f \Ou^\mu \\
-\frac {1}{4} &&\lambda_2\frac{1}{{\Lu}^{\dU}}\bar{\psi}i(\g_\mu{D_\nu}
+\g_\nu {D_\mu})\psi \Ou^{\mu\nu}  
\end{eqnarray}
\end{center}
and evaluate the process $e^+e^-\to f \bar{f}$ at ILC with center-of-mass
energy of 500 GeV. The $D_\mu$ stands for the
usual SM covariant derivative, while $\psi$ stands for the SM fermion
doublet or singlet. The $f$ is a SM fermion and $\lambda_i$'s are the
dimensionless effective couplings obtained after the imposition of
matching conditions on the non-trivial scale invariant sector below the
scale $\Lu$ \cite{kingman2}. The $\dU$ stands for the scaling dimension of
the unparticle operator $\Ou$. For simplicity, we have restricted ourselves
to $\lambda_i = 1$ for all $i=0,1,2$.
\begin{figure}[t!]
\vspace*{-0mm}
\rotatebox{0}{\includegraphics[height=0.3\textwidth]{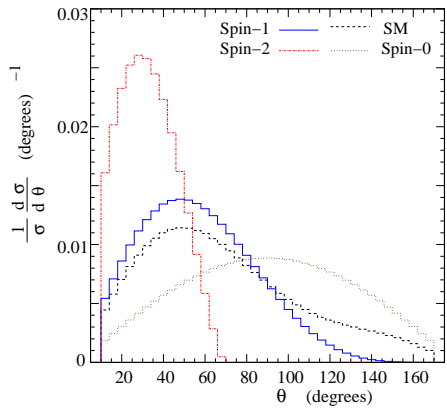}}
\vspace*{-0mm}
\caption{ The normalized differential cross-section plotted against the 
scattering angle $\theta$ for the final state $\mu$ in the scattering 
process $e^+e^-\to \mu^+ \mu^-$. The initial beams are longitudinally 
polarized with $P_L^{e^-}=0.8$ and $P_L^{e^+}=0.6$.}
\label{scatmu}
\end{figure}

We now take up each of the different cases separately. One of the very
distinct features that isolate the unparticles is the complex phase that
appears for the s-channel propagator. The virtual propagator for the
unparticles was found to have the following general form
\cite{georgi2,kingman1}
\begin{equation} \label{propagator}
\Delta_F^{\mathcal{U}} (P^2) = Z_{\dU} (-P^2)^{\dU - 2} \Pi_{\mathcal{U}}
\end{equation}
where 
$$
Z_{\dU} = \frac{1}{2\sin(\dU \pi)}
\frac{16\pi^{5/2}\G(\dU + \frac{1}{2})}{(2\pi)^{2\dU} \G(\dU-1)\G(2\dU)}
$$
and $\Pi_{\mathcal{U}}$ represent the spin structures. The detailed expressions for
them can be found in Ref.\cite{kingman2}. The interesting thing to note is
that for $P^2 > 0$, the complex function $(-P^2)^{\dU - 2}$ needs a branch
cut. This results in a complex phase to be associated for an s-channel
exchange of unparticles. This gives interesting effects on the
cross-sections for physical process due to its interference effects. In
this work we take up this feature of the unparticle propagator and try to
analyze the effects it can have on the azimuthal distribution of the final
state particles.

The general expression for the differential cross-section of scattering process 
$e^+e^-\to f \bar{f}$ can be written down as 
\begin{equation} \label{sigma}
\frac{d\sigma}{d\Omega} = \frac{1}{16\pi^2} \frac{\mathcal{F}}{2s^3} 
                           |{M_{SM+\mathcal{U}}}|^2
\end{equation}
where $\mathcal{F}=\lambda^{1/2} (s,m_f,m_f)$ and $d\Omega = \sin\theta
d\theta d\phi$. The spin-averaged amplitudes for the SM and the
different unparticles contributing in the virtual exchange is given by 
$|{M_{SM+\mathcal{U}}}|^2$ and is listed individually in
the appendix. We give the formulae for both longitudinal and transverse
beam polarizations. 
\begin{figure}[t!] 
\vspace*{-0mm}
\rotatebox{0}{\includegraphics[height=0.3\textwidth]{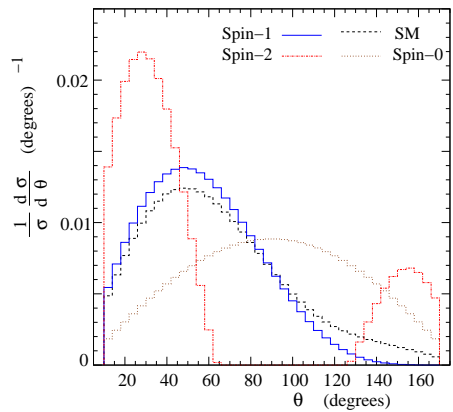}}
\vspace*{-0mm}
\caption{ The normalized differential cross-section plotted against the 
scattering angle $\theta$ for the final state $b$ quark in the scattering 
process $e^+e^-\to b\bar{b}$. The initial beams are longitudinally
polarized with $P_L^{e^-}=0.8$ and $P_L^{e^+}=0.6$.}
\label{scatb}
\end{figure}
\begin{figure}[htb]
\vspace*{-0mm}
\rotatebox{0}{\includegraphics[height=0.3\textwidth]{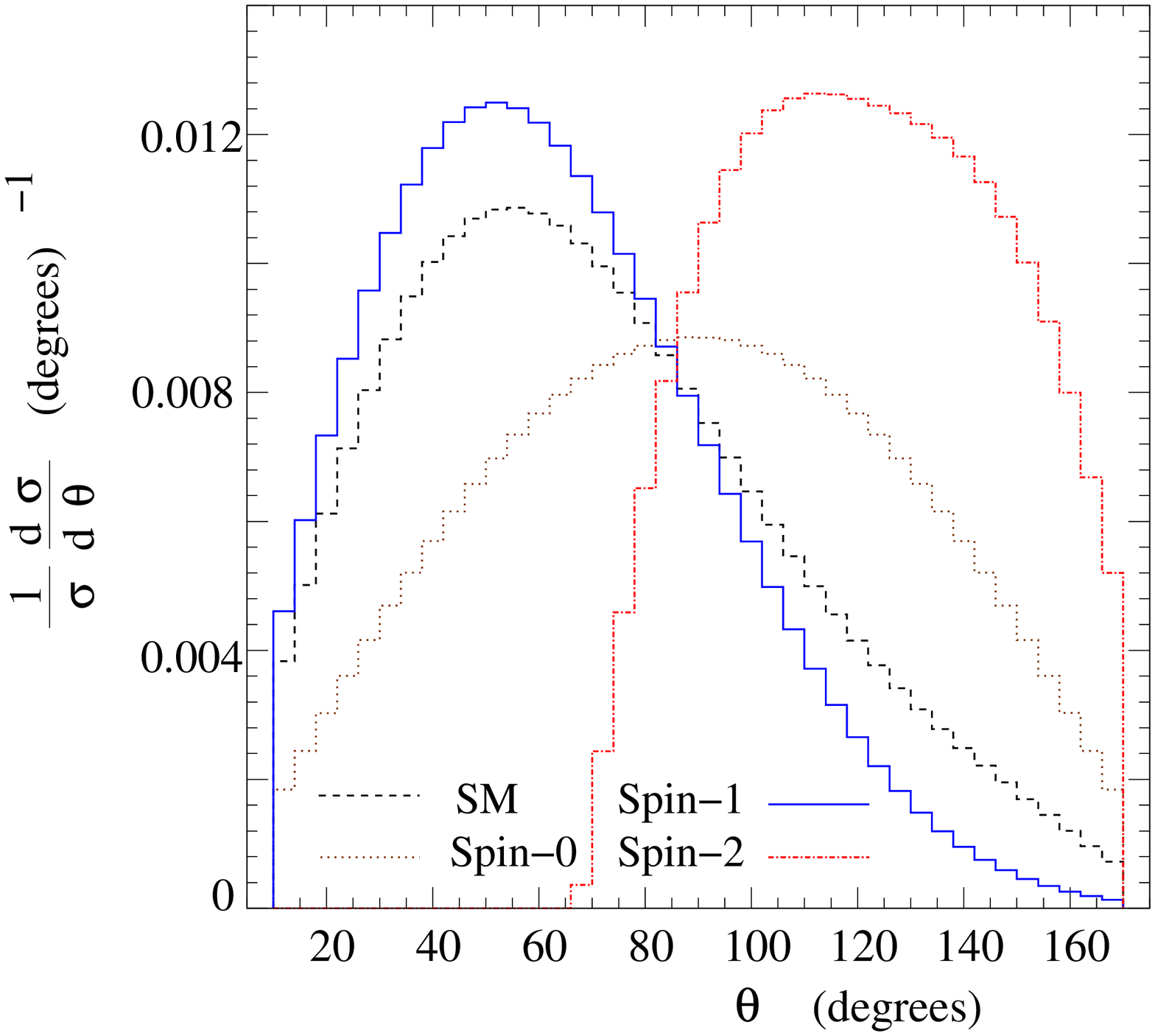}}
\vspace*{-0mm}
\caption{ The normalized differential cross-section plotted against the 
scattering angle $\theta$ for the final state $t$ quark in the scattering 
process $e^+e^-\to t\bar{t}$. The initial beams are longitudinally
polarized with $P_L^{e^-}=0.8$ and $P_L^{e^+}=0.6$.}
\label{scatt}
\end{figure}
Throughout our analysis, our choice for the free parameters
of the unparticle sector are as follows
\begin{eqnarray}
\Lu = 10&& ~{\rm TeV},~\dU=1.5~ (spin-0~{\rm and}~spin-1) \\
\Lu =  1&& ~{\rm TeV},~\dU=1.3~ (spin-2) 
\end{eqnarray} 
The above choice gives a significant number of new physics events for the 
different spin unparticles for an integrated luminosity in the range of 
$\mathcal{L}=10-100~fb^{-1}$.  

In Figs.\ref{scatmu},\ref{scatb} and \ref{scatt}, we plot the 
normalized differential cross-sections against the scattering angle for the
final state fermions. One can use longitudinally polarized beams which 
help in suppressing the SM background in the analysis for unparticles with
different spin options. Since our results are 
quite general, provided we have significant cross-section for new physics, we
just consider one case where the initial beam polarization is 
$P_L^{e^-}=0.8$ and $P_L^{e^+}=0.6$. Thus we have preferred to show the 
normalized distributions only. The notation we follow for the cross-sections 
plotted for different spin cases of the unparticles, is simply the excess 
over the SM cross-section. The excess cross-section we get by subtracting 
the SM cross-section from the total cross-section (obtained by taking the 
direct and 
\begin{figure}[htb]
\vspace*{-0mm}
\rotatebox{0}{\includegraphics[height=0.3\textwidth]{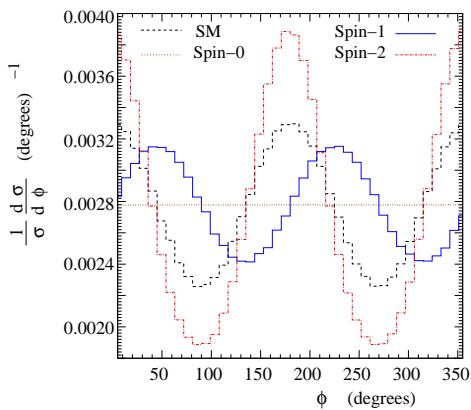}}
\vspace*{-0mm}
\caption{ The normalized differential cross-section plotted against the 
azimuthal angle $\phi$ for the final state $\mu$ in the scattering 
process $e^+e^-\to \mu^+ \mu^-$. The initial beams are transversely
polarized with $P_T^{e^-}= -0.8$ and $P_T^{e^+}=-0.6$.}
\label{azimmu}
\end{figure}
interference terms in the matrix amplitudes listed in the appendix) is 
used to plot the normalized distributions for the different spin exchanges of
the unparticles. One can see that for the massless fermions in the final state
i.e. $\mu$, the scattering angle distribution is helpful in differentiating the 
case of scalar exchange quite clearly. The tensor exchange does show a 
characteristic difference with a very pronounced peak at low $\theta$. The 
exchange of spin-1 unparticle will however show a similar profile to the
SM distribution. The clear dependence of mass of the final state fermions on 
the production cross-section also effects the scattering angle distribution for
the spin-2 exchange of the unparticles. This can be seen in the angular 
distributions plotted in Fig.\ref{scatb} and Fig.\ref{scatt} for the $b\bar{b}$
and $t\bar{t}$ final states, particularly for the top quarks in the final
state. However one needs to keep in mind that the heavy top quark will decay 
and the efficacy of this distinction would clearly be on how well the parent 
particle gets reconstructed from its decay products. Note here
that although the complex phase in the s-channel has a nontrivial effect on the
production cross-section as shown in earlier works 
\cite{georgi2,kingman1,kingman2}, its effects on the scattering angle 
distribution are minimal. 

At the next generation linear collider which would 
run for one or two fixed center-of-mass energies, it would be very 
difficult to see the effects of the complex phase on the cross-section. So
we should try to identify a variable which is susceptible to the complex phase. 
The most simple variable which can have a nontrivial dependence is the 
azimuthal angle ($\phi$). The $\phi$-dependent terms in the 
spin-averaged matrix amplitude square for the scattering process 
$e^+e^-\to f\bar{f}$ can be accessible if the initial beams
are transversely polarized \cite{rizzo}. The azimuthal angle is 
defined by the directions of the $e^\pm$ transverse polarization and the 
plane of the momenta of the outgoing fermions in the $e^+e^-\to f\bar{f}$ 
process. Writing the various matrix amplitudes for the SM and the unparticle 
exchanges in the s-channel with transversely polarized beams, one realizes 
that the complex phase gives a non vanishing 
\begin{figure}[t!]
\vspace*{-0mm}
\rotatebox{0}{\includegraphics[height=0.3\textwidth]{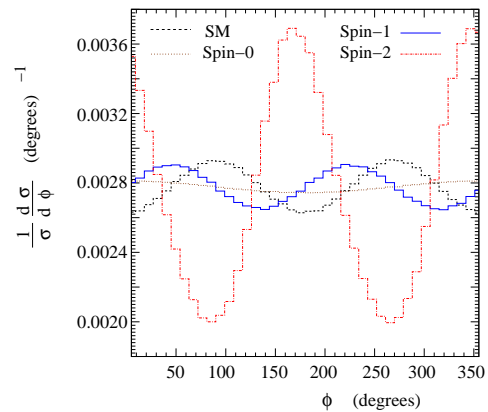}}
\vspace*{-0mm}
\caption{ The normalized differential cross-section plotted against the 
azimuthal angle $\phi$ for the final state $b$ quark in the scattering 
process $e^+e^-\to b\bar{b}$. The initial beams are transversely
polarized with $P_T^{e^-}= -0.8$ and $P_T^{e^+}=-0.6$.}
\label{azimb}
\end{figure}
imaginary component in the interference terms which are proportional to the 
azimuthal angle and contributes by interfering with the imaginary part of
the $Z$-boson propagator. 
Once integrated over the azimuthal angle, the transverse 
polarizations do not affect the total cross-section for the SM or
with the spin-1 and spin-2 exchange of the unparticles \cite{hikasa}. 
But one finds an imprint of the transversely polarized beams in the 
azimuthal angle distribution of the final state fermions. We try to exploit 
this fact in exploring the effect of the complex phase due to the s-channel 
unparticle propagator in the simplest scattering process at the linear 
collider. 
\begin{figure}[htb]
\vspace*{-0mm}
\rotatebox{0}{\includegraphics[height=0.3\textwidth]{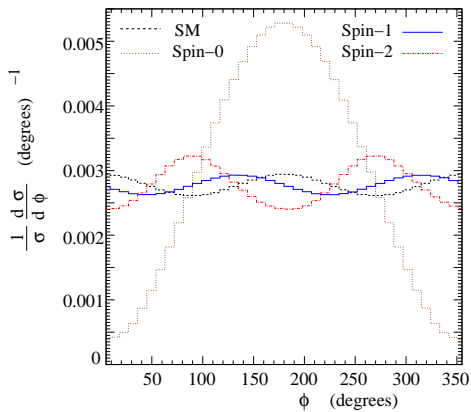}}
\vspace*{-0mm}
\caption{ The normalized differential cross-section plotted against the 
azimuthal angle $\phi$ for the final state $t$ quark in the scattering 
process $e^+e^-\to t\bar{t}$. The initial beams are transversely
polarized with $P_T^{e^-}= -0.8$ and $P_T^{e^+}=-0.6$.}
\label{azimt}
\end{figure}

In Figs.\ref{azimmu},\ref{azimb} and \ref{azimt} we plot the normalized 
differential cross-section against the azimuthal angle for the transversely 
polarized initial beams, for the three different final states {\it viz.} 
$\mu^+\mu^-, b\bar{b}$ and $t\bar{t}$ respectively. One finds that for the
spin-0 unparticle exchange, the azimuthal angle dependence is proportional 
to the mass of the final state fermions. Hence in Fig.\ref{azimmu}, we find 
that for the scalar exchange there is no azimuthal dependence, as the 
interference terms with the photon and the $Z$-boson vanish simultaneously for
vanishing fermion mass. However, the other unparticle exchanges do show a non 
trivial dependence. The most interesting thing to notice is the case for the
spin-1 exchange of the unparticle. It shows a clear phase difference of $\pi/4$
with the SM distribution. This is purely due to the complex phase, as the 
choice of $\dU$ is seen to control its distribution. 
The spin-2 exchange is seen to behave like a pure photon exchange 
distribution, with no $\g_5$ dependence. To explore the role of mass of the 
final state fermions in the distributions, we plot similar distributions 
for massive fermions in the final state. The choice for the input scale 
$\Lu$ and the scaling dimension $\dU$ remain the same as before. 
In Fig.\ref{azimb} we show the azimuthal dependence for the final state 
b-quarks. One marked difference is in the SM distribution itself. This can 
be understood as follows. The photon exchange and the $Z$-boson exchange 
diagrams have relatively different distributions. For the 
$\mu^+\mu^-$ final states, it is the photon exchange which dominates the 
distribution while it becomes just the opposite for the $b\bar{b}$ final 
states due to the smaller electromagnetic coupling of the $b$-quark with 
the photon. The spin-1 unparticle exchange also behaves like
$Z$-exchange. However, 
the tensor exchange is not affected by this. This shows that the 
$b\bar{b}$ final states can be easily used for identifying the
spin-2 exchange of unparticles. We however must point out that the complex 
phase does not play a role in the azimuthal distribution when a spin-2 
unparticle is exchanged. The scalar unparticle exchange shows a slight 
deviation from the flat distribution which is because the small mass of the 
$b$-quark gives non-vanishing interference terms with the SM diagrams. To
really highlight the effect of mass on the azimuthal distribution, we consider
the top quark in the final state and show the corresponding distribution in
Fig.\ref{azimt}. The most striking change is for the scalar unparticle 
exchange, which was expected since the interference terms were proportional
to the mass of the final state fermions. But it is also worth noting the 
effect of the mass on the distribution for the spin-2 exchange of unparticles. 
\begin{figure}[htb]
\vspace*{-0mm}
\rotatebox{0}{\includegraphics[height=0.3\textwidth]{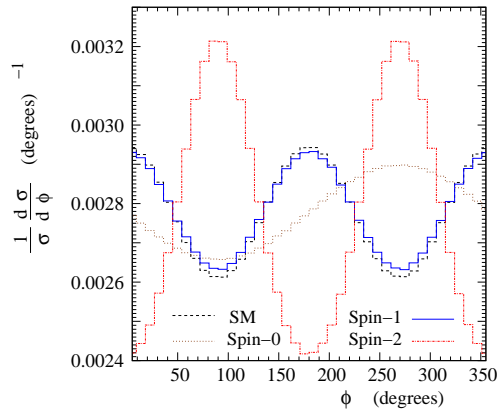}}
\vspace*{-0mm}
\caption{ The normalized differential cross-section plotted against the 
azimuthal angle $\phi$ for the final state $t$ quark in the scattering 
process $e^+e^-\to t\bar{t}$. We set the complex phase in the propagator
of the unparticles to zero to compare with its non-zero effects. The initial 
beams are transversely polarized with $P_T^{e^-}= -0.8$ and $P_T^{e^+}=-0.6$.}
\label{azimt0}
\end{figure}

In order to show the effect of the complex phase associated with the s-channel
unparticle propagator on the azimuthal distribution, we consider the 
process $e^+e^-\to t\bar{t}$ and set the complex phase to zero. We plot the
corresponding distribution in Fig.\ref{azimt0}. As expected the phase 
difference between the spin-1 unparticle exchange and SM vanishes, and even
the substantially modified curve for the scalar unparticle exchange 
suggests the non-trivial dependence, which the complex phase has on its 
distribution. We however point out that when setting the complex phase to
zero, the scaling dimension $\dU$ and the unparticle scale $\Lu$ has been 
taken as 
\begin{eqnarray*} \label{zerophase}
\Lu = 1&& ~{\rm TeV},~\dU=1.4~~(spin-0) \\
\Lu = 2&& ~{\rm TeV},~\dU=1.4~~(spin-1) \\
\Lu = 1&& ~{\rm TeV},~\dU=1.15~ (spin-2) 
\end{eqnarray*} 
so as to have enough significant cross-sections for the different cases.

\begin{figure}[htb]
\vspace*{-0mm}
\rotatebox{0}{\includegraphics[height=0.3\textwidth]{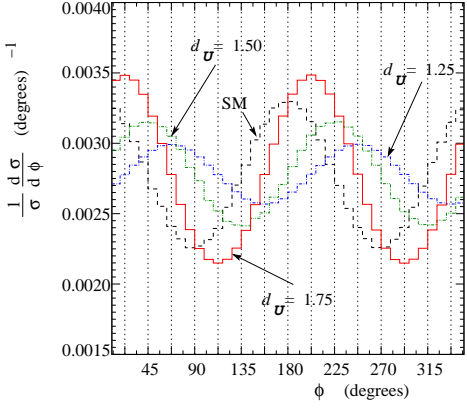}}
\vspace*{-0mm}
\caption{ The normalized differential cross-section plotted against the 
azimuthal angle $\phi$ for the final state $\mu$ in the scattering 
process $e^+e^-\to \mu^+\mu^-$. We consider the spin-1 unparticle contribution
in the s-channel and compare with the SM distribution. The initial 
beams are transversely polarized with $P_T^{e^-}= -0.8$ and $P_T^{e^+}=-0.6$.}
\label{phase}
\end{figure}
As pointed out earlier we find that for the spin-1 unparticle exchange, 
one can get a very strong estimate of the scaling dimension $\dU$ in an 
independent way. This can be done by just looking at the azimuthal 
distribution of the final state fermion and compare it with the pure SM 
distribution. To highlight this, we plot the normalised distributions for 
different values of $\dU$ for a spin-1 unparticle exchange in Fig.\ref{phase}. 
One finds that for $\Delta\dU=0.25$, the corresponding phase difference with 
the SM is $\pi/8$. This gives a very clean measure of the scaling 
dimension $\dU$ by just studying the azimuthal distribution of the fermion 
in the final state.
 
Thus we find that a simple analysis of the scattering angle and azimuthal 
dependence of the normalized differential cross-section for the simple 
minded scattering process $e^+e^-\to f\bar{f}$ at linear colliders with 
polarized beams can very effectively identify the effects of unparticle 
exchanges. By analyzing the above distributions at a 500 GeV linear 
electron-positron collider we find that one can easily exploit the 
azimuthal distribution to distinguish the spin of the exchanged unparticle 
by using transversely polarized initial beams, complemented by the scattering 
angle distributions. We must point out that one can have a non trivial
azimuthal dependence in supersymmetric theories which do not conserve 
R-parity, where one can consider scalar exchange in the propagator \cite{rmg}. 
However, the contributions to  the process $e^+e^- \to f \bar{f}$ would be
through the exchange of squarks and sleptons. The angular distributions would
be crucial and very effective in distinguishing the scalar exchanges of the 
supersymmetric particles from that of the spin-1 or spin-2 unparticle exchanges.To distinguish the effect of scalar
unparticles from supersymmetric contributions, the azimuthal distribution will 
still be a useful tool as the distributions with the complex phase differs
from the generic scalar exchanges, as shown in Fig.\ref{azimt0}. 

We also find an interesting fact that the spin-1 unparticle exchange 
shows a direct sensitivity to the choice the scaling dimension $\dU$ in the 
form of the phase difference with the SM distribution in $\phi$, 
and hence can provide for a direct measure of the scaling dimension, 
irrespective of the scale $\Lu$. This is an extraordinary result, bearing in 
mind that the complex phase associated with the s-channel exchange is a 
unique property of the unparticle theory. One can also consider different 
asymmetries using the dependence of the azimuthal angle to explore the 
effect of the complex phase in the unparticle propagator with time-like
momenta. Thus we see that with the different
nature of polarizations at the ILC, the unparticle sector of nature, if it
exists and is observable cannot escape our attention and will be found.

{\em Acknowledgments} 

SKR would like to thank S. Rindani and S. Roy for useful discussion. 
The authors gratefully acknowledge support from the Academy of Finland 
(project number 115032).

\begin{widetext}
{\bf \Large Appendix:}
\vspace*{5mm}

The amplitude with longitudinally polarized beams 
($P_L^{e^-},P_L^{e^+}$) for the SM and unparticle exchange can be written 
down as

\begin{eqnarray*}
M_{\g\g}^2 &=& Q_e^2 Q_f^2 e^4 \frac{N_c^f (1 - P_L^{e^-} P_L^{e^+})}{s^2} 
              \left[4u^2+4su+2s^2+4m_f^4-8um_f^2\right]\\
M_{\g Z}^2 &=& \frac{Q_e Q_f e^4}{16 s_W^2 c_W^2} 
\frac{N_c^f \left[(1 - P_L^{e^-} P_L^{e^+})e_V f_V
+ (P_L^{e^-}-P_L^{e^+})e_A f_V\right]}
{s~(s-M_Z^2+i\G_ZM_Z)} \left[(4u^2+4su+2s^2+4m_f^4-8um_f^2)\right] \\
&& - \frac{Q_e Q_f e^4}{16 s_W^2 c_W^2} 
\frac{N_c^f \left[(1 - P_L^{e^-} P_L^{e^+})e_A f_A
+ (P_L^{e^-}-P_L^{e^+})e_V f_A\right]}
{s~(s-M_Z^2+i\G_ZM_Z)} \left[(2s^2+4su-4sm_f^2)\right] \\
M_{Z Z}^2 &=&  \frac{e^4}{256 s_W^4 c_W^4}
\frac{N_c^f (f_V^2+f_A^2)\left[(1 - P_L^{e^-} P_L^{e^+})(e_V^2+e_A^2)
+2(P_L^{e^-}-P_L^{e^+})e_V e_A \right]}
{(s-M_Z^2)^2+(\G_ZM_Z)^2} \left[(4u^2+4su+2s^2+4m_f^4-8um_f^2)\right] \\
&& - \frac{e^4}{256 s_W^4 c_W^4}
\frac{N_c^f \left[4(1 - P_L^{e^-} P_L^{e^+})\right]} {(s-M_Z^2)^2+(\G_ZM_Z)^2} 
\left[e_V e_A f_V f_A (2s^2+4su-4sm_f^2) 
+ 2 s m_f^2 f_A^2 (e_V^2+e_A^2)\right] \\
&& - \frac{e^4}{256 s_W^4 c_W^4} 
\frac{N_c^f \left[2(P_L^{e^-}- P_L^{e^+})\right]}{(s-M_Z^2)^2+(\G_ZM_Z)^2} 
\left[f_V f_A (e_V^2+e_A^2) (2s^2+4su-4sm_f^2)\right] \\
M_{SS}^2 &=& \frac{\lambda_0^4 |\Delta_F (-s)|^2}{\Lu^{4\dU-4}} 
\frac{N_c^f (1 + P_L^{e^-} P_L^{e^+})}{2} (2s^2 - 5 s m_f^2 ), 
~~~M_{\g S}^2 = 0, ~~~M_{ZS}^2 = 0 \\
M_{VV}^2 &=& \frac{\lambda_1^4 |\Delta_F (-s)|^2}{\Lu^{4\dU-4}}
16 N_c^f ( 1 - P_L^{e^-} P_L^{e^+} + P_L^{e^-} - P_L^{e^+}) 
(u^2 - 2 u m_f^2 + m_f^4) \\
M_{\g V}^2 &=& \frac{Q_eQ_fe^2\lambda_1^2\Delta_F (-s)}{s \Lu^{2\dU-2}}
4 N_c^f ( 1 - P_L^{e^-} P_L^{e^+} + P_L^{e^-} - P_L^{e^+}) 
(u^2 - 2um_f^2 + 2sm_f^2 + m_f^4) \\
M_{Z V}^2 &=& \frac{4 N_c^f (e_V+e_A) e^2\lambda_1^2\Delta_F (-s)}
{16 (s-M_Z^2 - i\G_Z M_Z) \Lu^{2\dU-2} s_W^2 c_W^2}
( 1 - P_L^{e^-} P_L^{e^+} + P_L^{e^-} - P_L^{e^+}) 
\left[ (f_V+f_A) (u^2 - 2 u m_f^2 + m_f^4) + (f_V-f_A) s m_f^2 \right] \\
M_{TT}^2 &=& \frac{\lambda_2^4 |\Delta_F (-s)|^2}{256 \Lu^{4\dU}}
N_c^f ( 1 - P_L^{e^-} P_L^{e^+}) \left[ 256 u^4 + 512 s u^3 + 336 s^2 u^2 
+ 80 s^3 u + 8 s^4 - 32 m_f^2(32 u^3 + 40 s u^2 13 s^2 u + s^3) \right. \\
&& \left. + 16 m_f^4 (96 u^2 + 64 su + 5 s^2) 
- 256 m_f^6 ( 4 u + s - m_f^2) \right] \\
M_{\g T}^2 &=& \frac{- Q_eQ_f e^2 \lambda_2^2 \Delta_F (-s)}{16 s \Lu^{2\dU}}
N_c^f ( 1 - P_L^{e^-} P_L^{e^+}) \left[ 32 u^3 + 48 su^2 + 24 s^2 u 
+  4 s^3 - 8 m_f^2(12 u^2 + 8 s u  + s^2) \right. \\
&& \left. + 16 m_f^4 (6 u + s - 2 m_f^2) \right] \\
M_{ZT}^2 &=& \frac{-e^2 \lambda_2^2 \Delta_F (-s) N_c^f}
{256(s-M_Z^2 - i\G_Z M_Z)\Lu^{2\dU} s_W^2 c_W^2} \\
&& \left[( 1 - P_L^{e^-} P_L^{e^+}) \left[ 4 e_V f_V \left\{( 8u^3 + 12 su^2 
+ 6 s^2u + s^3) - 2 m_f^2(12u^2+8su+s^2) 
+ 4 m_f^4(6u+s-2m_f^2)\right\}  \right.\right.\\
&& \left.\left. - 4 e_A f_A \left\{(6 su^2+6s^2u+s^3) 
- 2 m_f^2(6su + 2s^2 - 3 s m_f^2)\right\}\right] \right. \\
&& + \left. (P_L^{e^-}- P_L^{e^+}) \left[ e_A f_V \left\{(8u^3 + 12 su^2 
+ 6 s^2u + s^3) - 2 m_f^2(12u^2+8su+s^2) + 4 m_f^4(6u+s-2m_f^2)\right\} 
\right.\right. \\
&& \left.\left. - 4 e_V f_A \left\{(6 su^2+6s^2u+s^3) 
- 2 m_f^2(6su + 2s^2 - 3 s m_f^2)\right\} \right]\right] .
\end{eqnarray*}
\vspace*{5mm}

Similarly the amplitude with transversely polarized beams 
($P_T^{e^-},P_T^{e^+}$) for the SM and unparticle exchange can be written 
down as

\begin{eqnarray*}
M_{\g\g}^2 &=& \frac{Q_e^2 Q_f^2 e^4}{s^2} 
\left [N_c^f (1 - P_T^{e^-} P_T^{e^+}) (4u^2+4su-8um_f^2+4m_f^4) 
+ N_c^f (2 s^2 + P_T^{e^-} P_T^{e^+} 
(2sm_f^2-2s^2) \sin^2\theta\cos^2\phi)\right]\\
M_{\g Z}^2 &=& \frac{Q_e Q_f e^4}{16 s_W^2 c_W^2} 
\frac{N_c^f} {s~(s-M_Z^2+i\G_ZM_Z)} 
\left[((1 - P_T^{e^-} P_T^{e^+})e_V f_V(4u^2+4su+4m_f^4-8um_f^2) 
- e_A f_A (2s^2 + 4su - 4s m_f^2) \right. \\
&& \left. + 2 e_V f_V s^2 + e_V f_V P_T^{e^-} P_T^{e^+} 
(2sm_f^2 - 2s^2) \sin^2\theta\cos^2\phi + i e_A f_V P_T^{e^-} P_T^{e^+} 
(2sm_f^2 - 2s^2) \sin^2\theta\sin\phi\cos\phi \right] \\
M_{Z Z}^2 &=&  \frac{e^4}{256 s_W^4 c_W^4}
\frac{N_c^f} {(s-M_Z^2)^2+(\G_ZM_Z)^2} 
\left[(1 - P_T^{e^-} P_T^{e^+})(e_V^2+e_A^2)(f_V^2+f_A^2)(4u^2+4su
+4m_f^4-8um_f^2)\right. \\
&& \left. + \left\{ f_A^2(e_V^2+e_A^2)(2s^2-8sm_f^2) 
+ 2 s^2 f_V^2 (e_V^2+e_A^2) 
- 8 e_V e_A f_V f_A (s^2 + 2su -2 s m_f^2) \right.\right. \\
&& \left.\left. + P_T^{e^-} P_T^{e^+}(e_V^2+e_A^2)(f_V^2+f_A^2)(2s^2-2 s m_f^2) 
\sin^2\theta\cos^2\phi \right\}\right] \\
M_{SS}^2 &=& \frac{\lambda_0^4 |\Delta_F (-s)|^2}{\Lu^{4\dU-4}} 
\frac{N_c^f (1 - P_T^{e^-} P_T^{e^+})}{2} (2s^2 - 5 s m_f^2 ) \\
M_{\g S}^2 &=& \frac{Q_e Q_f e^2 \lambda_0^2 \Delta_F (-s)}{2 s \Lu^{2\dU-2}}
 \left[ - 4 i N_c^f (P_T^{e^-} - P_T^{e^+})\sqrt{s - 4m_f^2}
s m_f\sin\theta \cos\phi\right] \\
M_{ZS}^2 &=&  \frac{e^2\lambda_1^2\Delta_F (-s)} 
{16 (s-M_Z^2 - i\G_Z M_Z) \Lu^{2\dU-2} s_W^2 c_W^2} 
\left[2N_c^f (P_T^{e^-} - P_T^{e^+})f_V \sqrt{s - 4m_f^2} s m_f \sin\theta 
(e_A \cos\phi - i e_V \sin\phi) \right]\\
M_{VV}^2 &=& \frac{\lambda_1^4 |\Delta_F (-s)|^2}{\Lu^{4\dU-4}}
 \left[ 16 N_c^f (u^2 - 2 u m_f^2 + m_f^4) \right] \\
M_{\g V}^2 &=& \frac{Q_e Q_f e^2\lambda_1^2\Delta_F (-s)}{s \Lu^{2\dU-2}}
N_c^f \left[ (1 - P_T^{e^-}P_T^{e^+}) (4u^2-8um_f^2+4m_f^4) + 4s m_f^2 
- 2 s P_T^{e^-}P_T^{e^+}(4u-(s-m_f^2)\sin^2\theta\cos\phi e^{i\phi}) \right]\\
M_{Z V}^2 &=& \frac{N_c^f (f_V+f_A) e^2\lambda_1^2\Delta_F (-s)}
{16 (s-M_Z^2 - i\G_Z M_Z) \Lu^{2\dU-2} s_W^2 c_W^2}
\left[ (e_V+e_A) (4u^2 - 8 u m_f^2 - 4 s m_f^2 + 4m_f^4 ) \right.  \\
&& \left. - (e_V-e_A) 
P_T^{e^-}P_T^{e^+} \left\{ (4u^2 - 8 u m_f^2 - 4 s u + 4m_f^4 ) 
+ (2 s^2 - 2 s m_f^2) \sin^2\theta \cos\phi e^{i\phi}\right\} \right] \\
M_{TT}^2 &=& \frac{\lambda_2^4 |\Delta_F (-s)|^2}{256 \Lu^{4\dU}}
N_c^f \left[ 256 (1 - P_T^{e^-} P_T^{e^+}) (u^4 + 2su^3 -4u^3 m_f^2 
- 5su^2m_f^2 + 6 u^2 m_f^4 + 4 su m_f^4 - 4u m_f^6 - s m_f^6 + m_f^8)
\right. \\
&& \left. + 8~(42 s^2 u^2 + 10 s^3 u + s^4 - 52 s^2 u m_f^2 - 4 s^3 m_f^2 
+ 10 s^2 m_f^4) - 8 P_T^{e^-} P_T^{e^+} ( 38 s^2 u^2 + 6 s^3 u - 44 s^2 u m_f^2
+ 6 s^2 m_f^4) \right. \\
&& \left. - 8 P_T^{e^-} P_T^{e^+} (16 s^2 u^2 + 16 s^3 u + 3s^4 -16 su^2 m_f^2
- 48 s^2u m_f^2 - 19 s^3 m_f^2 + 32 su m_f^4 + 32 s^2 m_f^4 - 16 s m_f^6) 
 \sin^2\theta \cos^2\phi \right] \\
M_{\g T}^2 &=& \frac{- Q_e Q_f e^2 \lambda_2^2 \Delta_F (-s)}{16 s \Lu^{2\dU}}
N_c^f \left [ 16 (1 - P_T^{e^-} P_T^{e^+}) (2 u^3 + 3 su^2 - 6 u^2 m_f^2
- 4 su m_f^2 + 6 u m_f^4 + s m_f^4 - 2 m_f^6) \right. \\
&& \left. + 24 s^2 u + 4 s^3 - 8 s^2 m_f^2  - 16 P_T^{e^-} P_T^{e^+} s^2u
- 8  P_T^{e^-} P_T^{e^+}(s^3 + 2s^2 u - 2su m_f^2 - 3s^2 m_f^2 
+ 2s m_f^4)\sin^2\theta\cos^2\phi \right] \\
M_{ZT}^2 &=& \frac{-e^2 \lambda_2^2 \Delta_F (-s)}
{256(s-M_Z^2 - i\G_Z M_Z)\Lu^{2\dU} s_W^2 c_W^2} 
N_c^f \left [ 16 (1 - P_T^{e^-} P_T^{e^+}) e_V f_V ( 2u^3 + 3 su^2 + s^2u
-6u^2 m_f^2 - 4su m_f^2 + 6u m_f^4 \right. \\
&& \left. + s m_f^4 - 2m_f^6) + 8 (P_T^{e^-} P_T^{e^+} - 3)e_A f_A 
(su^2 + s^2u - 2 su m_f^2 + s m_f^4) - 4 e_A f_A s^2 (s - 4 m_f^2) \right. \\
&& \left. + 4 e_V f_V s^2 (2u + s - 2m_f^2) + 4 P_T^{e^-} P_T^{e^+} 
\left\{s^2f_A(s - m_f^2)(e_A \cos\phi - ie_V\sin\phi) \right.\right. \\
&& \left. \left. - 2s f_V(2 su + s^2 - 2 u m_f^2 - 3 s m_f^2 + 2 m_f^4) 
(e_V\cos\phi - i e_A \sin\phi)\right\} \sin^2\theta\cos\phi\right ].
\end{eqnarray*}
\vspace*{3mm}

In the above formulae for the matrix amplitude squares, $m_f$ stands for the
mass of the final state fermions, $N_c^f$ stands for the color factor which 
equals 1 for leptons and 3 for quarks. The $s$ and $u$ represent the usual 
Mandelstam variables while $M_Z$ and $\G_Z$ stand for the mass and total 
width of the $Z$-boson. The variables $\theta$ and $\phi$ are the scattering 
angle and the azimuthal angles for the final state fermions respectively, while
$e_V = 2 T_3^e - 4 Q_e s_W^2,~ e_A= -2 T_3^e$ and 
$f_V = 2 T_3^f - 4 Q_f s_W^2,~ f_A= -2 T_3^f$. We have absorbed the spin 
structures while calculating the amplitude squares and hence 
$$
\Delta(-s) = Z_{\dU} s^{\dU-2} e^{-i\dU\pi} 
$$
The spin averaged matrix amplitude square is given by the sum written in the
form
$$
\sum_{i,j}^{i=j}M_{ij}^2+2\mathcal{R}e\left(\sum_{i,j}^{i\ne j}M_{ij}^2\right)
$$
where the $i,j = \g,Z,S(V,T)$ stand for the particles exchanged in the 
s-channel. The $S,V$ and $T$ stand for the scalar, vector and tensor unparticle
respectively.
\end{widetext}


\end{document}